\documentclass[manuscript,screen,nonacm]{acmart}

\usepackage{algorithmic}
\usepackage{quantikz}

\usetikzlibrary{backgrounds,fit} 

\newcommand{\DFT}{\mathrm{DFT}}
\newcommand{\W}[1]{\omega_{#1}}          
\newcommand{\I}[1]{I_{#1}}               
\newcommand{\Lperm}[2]{L^{#1}_{#2}}      
\newcommand{\Twid}[2]{T^{#1}_{#2}}       
\newcommand{\Twidtil}[2]{{\tilde{T}}^{#1}_{#2}}       
\newcommand{\Eone}{E_{1}}                
\newcommand{\Etwo}{E_{2}}                
\newcommand{\R}[1]{R_{#1}}               

\newlength{\imgwidth}
\title{Not Your Usual FFT: QFT→FFT via Classical Quantum-Circuit Simulation}

\author{Stefano Markidis}
\correspondingauthor
\email{markidis@kth.se}
\orcid{0000-0003-0639-0639}
\author{Gilbert Netzer}
\email{noname@kth.se}
\orcid{0000-0002-9479-7393}
\author{Luca Pennati}
\email{pennati@kth.se}
\orcid{0009-0009-4901-1716}
\author{Frej Larssen}
\email{flarssen@kth.se}
\orcid{0009-0002-8574-2021}
\author{Ivy Peng}
\email{bopeng@kth.se}
\orcid{0000-0003-4158-3583}
\affiliation{%
  \institution{KTH Royal Institute of Technology}
  \city{Stockholm}
  \country{Sweden}
}

\renewcommand{\shortauthors}{S. Markidis et al.}

\begin{abstract}
  We introduce QFT→FFT, a family of HPC FFT libraries that compute the discrete Fourier transform by executing a quantum Fourier transform (QFT) circuit on classical quantum computer simulators. Input arrays are mapped directly to state amplitudes with explicit normalization/indexing, making QFT a drop-in replacement for FFT primitives. A backend-agnostic planner builds a fused-gate schedule and memory layout adapters to increase arithmetic intensity and reduce memory data movement. We implement this design on top of Google’s C++ \texttt{qsim} and evaluate OpenMP, AVX, and CUDA backends. On an AMD EPYC Zen2 processor, our AVX performance is on par with that of multithreaded FFTW, utilizing 64 threads. On an NVIDIA A100, the CUDA backend achieves more than $4\times$ lower time than both AVX and FFTW on AMD EPYC Zen2 at larger sizes. We also employ an approximate QFT (AQFT) that truncates small-angle controlled rotations beyond a cutoff $k$, reducing circuit depth and runtime while preserving accuracy.
\end{abstract}

\begin{document}

\maketitle

\section{Introduction}
Discrete Fourier Transforms (DFTs) are a critical computational primitive in many applications, ranging from performing spectral analysis of a given signal to finding spectral solutions of partial differential equations~\cite{alessio2015digital}. Almost all HPC DFT libraries implement the so-called Fast Fourier Transform (FFT) algorithms~\cite{van1992computational}, which reduce the computational cost of the DFT calculation from $O(N^2)$ of the naive DFT algorithm to $O(N\log N)$, where $N$ is the size of the input array of complex numbers. HPC libraries, such as the Fastest Fourier Transform in the West (FFTW)~\cite{frigo1998fftw}, Intel MKL FFT~\cite{wang2014intel}, and Spiral FFT \cite{franchetti2006fft}, are some of the de-facto options both in the algorithmic choice and HPC implementations. New FFT implementations vary for different hardware architectures, including adapting vectorized instructions~\cite{franchetti2005efficient}, using tensor cores~\cite{li2021tcfft}, and employing more advanced compiler technologies~\cite{kopcke2019generating} leading to interesting variations of a theme. Given the optimality of the approach, major recent novelty in other DFT technologies are observed to focus on improved implementations targeting new hardware. 

On the other hand, quantum computing is arising as a new computational paradigm, bringing in new concepts and a different way to perform computation~\cite{markidis2024quantum}. A central primitive is the quantum Fourier transform (QFT), which, on a quantum device, implements a DFT with $O(\log^2 (N))$ gates, an asymptotic improvement over the classical $O(N\log N)$ FFT when gate cost is the metric. As opposed to classical FFT, in quantum computing, QFT is not meant to be run as a standalone DFT primitive for spectral analysis, as the cost of the signal encoding and the number of measurements to reconstruct the quantum state (the so-called I/O bottleneck~\cite{hoefler2023disentangling}) would negate the computational improvement of the QFT. Instead, QFT is meant to be executed as part of a larger application, such as Shor's algorithm, Quantum Phase Estimation (QPE)~\cite{nielsen2010quantum}, and Quantum Fourier Neural Operator neural networks~\cite{jain2024quantum}. Indeed, the quantum advantage of the most famous Shor's algorithm relies critically on the usage of the QFT~\cite{lin2014shor}.

While in quantum algorithms, the QFT is rarely used as a stand-alone transform, in classical HPC, state-vector simulators expose the full amplitude array after each gate application. This makes it natural to treat the QFT as a computational primitive for the DFT: load the input array as complex amplitudes, execute a QFT circuit, and read back the transformed amplitudes. This paper investigates the idea, QFT→FFT, as an alternative path to HPC DFTs on classical hardware. QFT→FFT executes entirely on classical HPC systems via quantum computer simulation. This differs fundamentally from running QFT on a quantum device. We do not face state-preparation or measurement overheads, but we incur classical memory sweeps inherent to classical simulator updates. Throughout, we therefore treat QFT→FFT as a classical HPC library that happens to be expressed as circuit execution, not as an experiment on quantum hardware. Although the QFT originates in a different community, its building blocks are familiar to HPC: $2\times2$ butterfly mixes, diagonal phase scales, and permutations. We show how to turn these into a planner-driven library that targets CPUs and GPUs, and how the approximate QFT (AQFT) can be obtained by truncating small controlled rotations.

For investigation and demonstration, we build our DFT library atop Google’s C++ state-vector simulator library \texttt{qsim}~\cite{isakov2021simulations}. We map the input array to a normalized state vector with explicit indexing, materialize the QFT as a circuit, and execute it on \texttt{qsim}’s CPU and GPU backends (multithreaded OpenMP, AVX, and CUDA). \texttt{qsim} performs gate fusion and inserts layout adapters to raise arithmetic intensity and reduce memory traffic. The resulting fused schedule is run by \texttt{qsim} as it is, without modifying simulator kernels or adding vendor-specific code. On an AMD EPYC Zen2, our AVX backend matches multithreaded FFTW across mid-to-large sizes, showing state-of-the-art parity between a QFT→FFT FFT library and a production FFT library, such as FFTW.

The contributions of this paper are the following:
\begin{itemize}
\item We define a QFT→FFT approach in which the FFT is realized by executing a QFT circuit as the core primitive. We specify the input array to amplitude mapping so the resulting operation reproduces standard DFT semantics and can serve as a drop-in FFT.
\item We implement the interface atop Google’s C++ \texttt{qsim} with multithreaded CPU, AVX, and CUDA backends. We show that the same high-level plan can target heterogeneous hardware without changing user codes.
\item We show that on an AMD EPYC Zen2, the AVX-enabled implementation is comparable to multithreaded FFTW and scales to 64 threads. On an NVIDIA A100, the CUDA backend achieves approximately $5\times$ speedup over the AVX backend and more than $4\times$ speedup over FFTW at larger problem sizes measured on the AMD EPYC CPU.
\item We show that truncating small-angle controlled rotations beyond a cutoff reduces circuit depth and runtime, leading to additional speedups on both CPU and GPU at the cost of reduced accuracy.
\end{itemize}

\section{Background}
In this section, we present the mathematical relationship between DFT, FFT, and QFT~\cite{camps2021quantum}. We then express it in terms of quantum circuits, and finally we delineate the computational complexity.\\

\noindent \textbf{From Naive DFT to FFT.} 
For presenting the different steps in going from naive DFT to FFT, we use factorization techniques, as presented in seminal works on the topics~\cite{van1992computational,franchetti2011fft,franchetti2003short,camps2021quantum}. We refer interested reader for a detailed explanation. Let $N$ be the transform size (or the size the input array) and $\W{N}=e^{-2\pi i/N}$. The (unnormalized) DFT matrix is $\DFT_N = [\omega_N^{k\ell}]_{0\le k,\ell<N}$. The size-2 DFT
\begin{equation}
\DFT_2 \equiv H  =\begin{bmatrix}1&1\\[2pt]1&-1\end{bmatrix},
\end{equation}
is called \textit{Hadamard transformation}. Applying $\DFT_{N}$ to a vector is a dense $N\times N$ matrix-vector (matvec) multiplication, leading to $O(N^2)$ complex flops. The FFT replaces the dense DFT matvec by a product of sparse/tensor factors. Writing $N=KM$, the Cooley-Tukey factorization is
\begin{equation}
\DFT_{N} = (\DFT_{K}\otimes \I{M})\;\Twid{KM}{M}\;(\I{K}\otimes \DFT_{M})\;\Lperm{KM}{K},
\label{dft_N}
\end{equation}
where the \emph{stride permutation} $\Lperm{KM}{K}: iM{+}j\mapsto jK{+}i$ regroups data and the \emph{twiddle} $\Twid{KM}{M}$ is a block-diagonal phase change. The rule is recursive. We apply it again to $\DFT_M$ and $\DFT_K$, and continue until size-$2$ kernels, producing $\log_2 N$ stages (for a radix-2 scheme). To calculate the computational complexity, let $T(N)$ be the arithmetic work. Since each stage consists of one butterfly bank (sparse $(\DFT_2\otimes I)$ or $(I\otimes \DFT_2)$ with $2N$ nonzeros), one twiddle (diagonal, $N$ nonzeros), and a permutation ($N$ nonzeros), the per-stage cost is $O(N)$, giving the recurrence
$$
\begin{aligned}
T(KM) &= K\,T(M) + M\,T(K) + c\,KM,\\
\text{so for } M=2:\quad T(N) &= c\,N + 2\,T(N/2)= O(N\log N).
\end{aligned}
$$

For demonstration, we focus on the factorization of $\DFT_{8}$, which corresponds to a two-level radix-2 factorization (first $8=4\cdot2$, then $4=2\cdot2$):
\begin{equation}
\DFT_{8}
=
\Big[(H \otimes \I{2})\;\Twid{4}{2}\;(\I{2}\otimes H)\;\Lperm{4}{2}\Big]\otimes \I{2}\;\;
\Twid{8}{2}\;\;(\I{4}\otimes  H)\;\;\Lperm{8}{4}.
\label{dft8}
\end{equation}
The operation sequence should be read right-to-left. First a stride change $\Lperm{8}{4}$, then a bank of $2\times 2$ butterflies $(\I{4}\otimes H)$ followed by a level-8 twiddle $\Twid{8}{2}$ and the factored $\DFT_4$ (tensored with $\I{2}$). \\

\noindent \textbf{From FFT to QFT.} We fix $M{=}2$, $K{=}N/2$ in the Cooley-Tukey rule, and group all $L$-factors \footnote{We do this by applying the formula
$A\otimes B = L_{K}^{KM} (B \otimes A) L_{M}^{KM}$, where 
$A$ is $K\times K$ and $B$ is $M \times M$\cite{franchetti2011fft} to move all $L$-factors to the left. $\tilde{T}$ is the resulting twiddle.
}
.
This leads to the radix-2 block form
\begin{equation}
    \begin{cases}
        \DFT_N &= \big(\text{product of $L$'s}\big) \times \DFT_N^* \;\\
        \DFT_N^* &= \big(\I{2} \otimes \DFT_{N/2}^*\big)\;{\Twidtil{N}{2}}\;\big(H \otimes \I{N/2}\big)
    \end{cases}.
\end{equation}

Note that the middle factor $\tilde{T}$ is a diagonal block separating two butterfly banks. The step from FFT to a QFT-style decomposition consists of investigating the diagonal twiddle and expressing it as a product of two-level Kronecker-sparse operators built from selectors and diagonals~\cite{camps2021quantum}. This can be achieved by defining the $2\times2$ projectors $E_1, E_2$ and the phase shift matrix $R_j$:
$$
\Eone=\operatorname{diag}\!\big(1,0\big),\qquad
\Etwo=\operatorname{diag}\!\big(0,1\big),\qquad
\R{j}=\operatorname{diag}\!\big(1,\W{2^{j}}\big).
$$
Intuitively, $\Eone/\Etwo$ selects whether a given binary digit of an index is $0$ or $1$ and acts as a selector.

%
%
A final step in moving from the FFT factorization to the QFT-style form is to break the diagonal $\tilde{T}$ into a short product of two-level (i.e., $4\times4$ diagonal) updates. If $b$ is the number of more-significant binary digits exposed at the stage (so the local diagonal has size $2^{b+1}$), then
\begin{equation}
\begin{split}
\Twidtil{2^{\,b+1}}{2}
&=
\prod_{i=1}^{b}
\Big[\, \big(\I{1}\otimes\I{i-1}\otimes\Eone+\R{i+1}\otimes\I{i-1}\otimes\Etwo \big)\otimes \I{b-i}\,\Big],\\
&\text{for } b=1,2,\dots,n-1.
\end{split}
\end{equation}
Each bracketed factor acts only on two $2\times2$ slots. It passes those entries unchanged when the selector is $\Eone$, and applies the diagonal $\R{i+1}=\mathrm{diag}(1,\omega_{2^{i+1}})$ when the selector is $\Etwo$.

Using these results, we can express $\DFT_{8}$ as a QFT. We start from the factorization, expressed in Eq.~\ref{dft8} that consists of three $H$-based butterfly banks, separated by diagonal blocks $T$ and stride permutations $L$. The first diagonal between the two leftmost banks is
$$
\Twidtil{4}{2}=\I{1}\otimes \Eone\;+\;\R{2}\otimes \Etwo.
$$
This is a single two-level update. Entries selected by $\Eone$ pass unchanged, while those selected by $\Etwo$ acquire the diagonal $\R{2}=\mathrm{diag}(1,\omega_{4})$. 

The second diagonal is larger, but it has the same structure. It factors as
\begin{align*}
\Twidtil{8}{2}
&=
\big(\I{1}\otimes \I{1}\otimes \Eone + \R{3}\otimes \I{1}\otimes \Etwo\big)\;
\big(\I{1}\otimes \Eone\otimes \I{1} + \R{2}\otimes \Etwo\otimes \I{1}\big).
\end{align*}
Altogether, $\DFT_{8}$ consists of three $H$-butterfly stages interleaved with one and then two two-level diagonal updates, while $\Lperm{8}{4}$ and $\Lperm{4}{2}$ handle the reindexing.

Relative to the standard FFT, only the diagonal changes in the QFT. The factor $T$ is refined into a short product of Kronecker-sparse two-level updates, which is the linear-algebraic core of the QFT-style decomposition. The butterfly and permutation structure remains unchanged. \\

\noindent \textbf{From Linear Algebra to Quantum Circuit Operators.}  A \emph{qubit} is a two-dimensional complex vector $|\psi\rangle=\alpha\,|0\rangle+\beta\,|1\rangle$ with $|\alpha|^2+|\beta|^2=1$. As a column vector, $|\psi\rangle=(\alpha,\beta)^{\!\top}\!\in\mathbb{C}^2$.  An $n$-qubit \emph{state vector} lives in $\big(\mathbb{C}^2\big)^{\otimes n}\cong\mathbb{C}^{2^n}$. We index its $N=2^n$ entries by binary strings of length $n$. A quantum gate $U$ is a unitary matrix that multiplies this state vector. A 1-wire gate $U\in\mathbb{C}^{2\times 2}$ acting on qubit $q_j$ applies the structured operator
$$
I_{2^{n-1-j}}\ \otimes\ U\ \otimes\ I_{2^{j}}\ \in\ \mathbb{C}^{2^n\times 2^n},
$$
In a state-vector quantum circuit simulator, it updates $2^{n-1}$ independent \emph{pairs} of amplitudes via the same $2\times2$ matvec. Two-wire gates similarly act on $2^{n-2}$ disjoint \emph{quartets} with a $4\times4$. 
For example, a butterfly bank $(H\otimes I)$ or $(I\otimes H)$ is applied across many pairs at once \cite{asaka2020quantum}. A controlled-$U$ on two wires (target, then control) is
$$
I_2\otimes E_1 \;+\; U\otimes E_2.  
$$
For QFT, the key controlled gate is the controlled phase $R_k$ with
$$
R_k=\operatorname{diag}\!\big(1,\ e^{-2\pi i/2^{k}}\big),\qquad 
\text{so}\quad \text{controlled-}R_k=E_1\otimes I_2+E_2\otimes R_k.
$$
This gate is a two-wire diagonal update. In this case, selected amplitudes are multiplied elementwise by the corresponding phase. These controlled-$R_k$ operators commute. This formulation exactly corresponds to the matrix identities used to factor the diagonal $\tilde{T}$ in our QFT-style derivation.

\begin{figure}[t]
\centering
\resizebox{0.5\columnwidth}{!}{%
\begin{quantikz}[row sep=0.25cm, column sep=0.40cm]
\lstick{$q_0$} & \qw & \qw        & \ctrl{2}   & \qw        & \ctrl{1}   & \gate{H} & \swap{2} &\qw  \\
\lstick{$q_1$} & \qw & \ctrl{1}   & \qw        & \gate{H}   & \gate{R_2} & \qw      & \qw & \qw      \\
\lstick{$q_2$} & \gate{H} & \gate{R_2} & \gate{R_3} \gategroup[wires=1,steps=1,style={draw,dashed,rounded corners,inner xsep=2pt,inner ysep=2pt}]{} & \qw & \qw & \qw & \targX{} & \qw 
\end{quantikz}%
}
\caption{Three-wire QFT for $\DFT_{8}$ (MSB $q_2$ at bottom, LSB $q_0$ at top). $H$ is the $2{\times}2$ DFT; $R_2$ and $R_3$ are controlled phases. Swaps implement the output bit-reversal $P=\prod_i L_i$ (equivalently, permute outputs by $P$). In AQFT with cutoff $k{=}1$, keep $R_2$ and drop $R_3$.}
\label{fig:qft3-qft-aqft}
\end{figure}
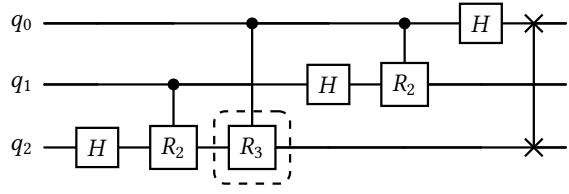

Let us now go back to the $\DFT_{8}$ example and check how it is constructed in quantum circuit formulation. The three $H$ operations realize the three $2\times2$ butterfly banks. The diagonal blocks $T$ factor into the controlled phases, and the SWAPs implement the same bit permutation as a product of stride operators $L$. The SWAP can be deferred to I/O or absorbed into layout. The final circuit layout is presented in Fig.~\ref{fig:qft3-qft-aqft}.

On a state-vector simulator, each $H$ updates $2^{n-1}=4$ amplitude pairs with a single $2\times2$ kernel, each controlled-$R_k$ is an elementwise diagonal scale on half the entries (masked by the control), and the final SWAPs are just index swaps.

An important concept for HPC optimization of quantum circuits is \textbf{gate fusion}. Consecutive gates acting on the same wires can be used by multiplying their small matrices first (e.g., $U_{\mathrm{fused}}=U_t\cdots U_2U_1$ with $U_i\in\mathbb{C}^{2\times2}$ or $4\times4$), then applying $U_{\mathrm{fused}}$ once to the state vector. Algorithmically, the gate fusion is a matrix-matrix multiply followed by a batched matvec. In addition, gates on disjoint wires can likewise be tensored (e.g., $H\otimes H$) and applied in a single pass. Finally, commuting diagonal factors (the controlled-$R_k$ terms) collapse to one diagonal by multiplying phases elementwise. 
Fusing is critical for performance, reduces memory accesses, and raises arithmetic intensity. Typically, in an HPC state-vector simulator, short fusion (2-4 gates) provides a good performance.\\

\noindent \textbf{Approximate QFT.} The QFT circuit for $N=2^n$ consists of $n$ Hadamards interleaved with diagonal, commuting controlled-$R_m$ gates whose angles shrink geometrically with the control--target distance.
From a matrix point of view, each stage’s diagonal $\tilde{T}$ is a product of $b$ two-level terms with phases $e^{-2\pi i/2^{i+1}}$. We define a range-$k$ AQFT by keeping only the first $k$ two-level terms per stage. This is equivalent to keeping only controlled rotations whose control-target distance is $\le k$ and dropping the rest. With no cutoff ($k=n-1$), the $n$-wire circuit has $n(n-1)/2$ controlled phases. In a state-vector simulator, truncation removes elementwise diagonal updates, directly reducing memory passes. On accuracy, every omitted factor has angle magnitude $\le 2\pi/2^{k+1}$. The resulting phase tail decays like $2^{-k}$, which is the standard intuition behind Coppersmith’s AQFT~\cite{coppersmith2002approximate}. For example, in the 3-wire QFT (Fig.~\ref{fig:qft3-qft-aqft}), $k=2$ is exact (keeps $R_2$ and $R_3$ on the bottom wire and $R_2$ on the middle), while $k=1$ keeps only nearest-neighbor rotations $R_2$ and drops $R_3$.\\

\noindent \textbf{Algorithmic Complexities.}  As the final part of this background section, we focus on computational complexity~\cite{musk2020comparison}, summarized in Tab.~\ref{tab:complexity}. The naive DFT is a dense $N\times N$ matvec. Therefore, it has $O(N^2)$ flops. The FFT factors the transform into $\log_2 N$ stages, each $O(N)$ (butterflies, diagonals, permutations), for an overall $O(N\log N)$ cost. For the QFT with $N=2^n$ on a quantum device, the circuit has $n$ Hadamards and $n(n-1)/2$ controlled phase rotations, i.e., $O(\log^2 (N))$ one-/two-qubit gates (depth $O(\log N)$ up to the optional bit reversal). On a classical state-vector simulator, each constant-size gate still sweeps $O(N)$ amplitudes (batched $2\times2$ or $4\times4$ updates), so the exact QFT costs $O(N\log^2 (N))$ flops, indeed slower than FFT. Note that the quantum advantage of QFT comes from performing the tensorial operation for $T$ across all the qubits in parallel, exploiting the \emph{principle of local operations}~\cite{ekert1998quantum}, while a state vector needs to sweep $O(N)$ amplitudes. The AQFT with cutoff $k$ keeps only controlled rotations of order $m\le k+1$, reducing the gate count to $O(k\log N)$ while leaving the $O(\log N)$ Hadamards unchanged. The simulator cost becomes $O(N k \log N)$. 
\begin{table}[t]
\centering
\small
\begin{tabular}{ll}
\toprule
\textbf{Algorithm} & \textbf{Asymptotic cost} \\
\midrule
Naive DFT (classical)              & $O(N^{2})$ flops \\
FFT (classical)      & $O(N \log N)$ flops \\
QFT (on quantum computer, exact)          & $O(\log^{2} N)$ gates \\
AQFT (on quantum computer, range-$k$)          & $O(k\log N)$ gates \\
QFT (state-vector simulator, exact) & $O(N \log^{2} N)$ flops \\
AQFT (state-vector simulator, range-$k$) & $O(N\,k\,\log N)$ flops \\
\bottomrule
\end{tabular}
\caption{Asymptotic computational cost for size-$N$ transforms. Classical rows report floating-point operations (flops); the quantum row reports one-/two-qubit gate counts with $N=2^n$. The simulator costs assume a state-vector model in which each constant-size gate sweeps $O(N)$ amplitudes.}
\label{tab:complexity}
\end{table}

\section{Methodology}
Our goal is to use the QFT as a DFT inside an HPC FFT workflow, using a quantum computer simulator. We proceed in four steps:
\begin{enumerate}
\item \emph{Input encoding.} Given a complex array $x$ of length $N$, we choose $n=\lceil \log_2 N\rceil$ qubits and allocate a $2^n$-element state vector $|\psi\rangle$. We fix and document a bit-indexing convention (endianness and bit order) so that index $k$ in the array unambiguously maps to the basis state $\lvert b_{n-1}\dots b_0\rangle$. Because quantum states must have unit $\ell_2$ norm, we normalize the simulator’s amplitudes. We compute a scalar \texttt{scale\_in} $=\lVert x \rVert_2$ (or a user-chosen equivalent), write $|\psi\rangle \leftarrow x/\texttt{scale\_in}$ under the chosen indexing, and record \texttt{scale\_in} as metadata. 
\item \emph{Circuit construction.} We build the standard $n$-qubit QFT according to Figure~\ref{fig:qft3-qft-aqft}. The final swaps are not implemented as gates but accounted for in the caller as an output index permutation.
\item \emph{State evolution.} The simulator applies the circuit gates in sequence to $\psi$; no measurement is performed, so the evolution is deterministic and preserves the state norm. The unitary used is the conventional, unit-norm QFT $U=\frac{1}{\sqrt{N}}DFT_N$. 

\item \emph{Output decoding.} Let $|\psi'\rangle$ be the final simulator state. To compare with classical FFT libraries under a specific scaling convention, we rescale $|\psi'\rangle$ back to \emph{FFT units} using the single scalar we tracked. With the unitary QFT above, $|\psi'\rangle \;=\; \frac{1}{\texttt{scale\_in}}\;\frac{1}{\sqrt{N}}\;DFT_N\,x$. To recover the unnormalized forward FFT result $DFT_N x$ we multiply by $\texttt{scale\_in}\sqrt{N}$, i.e., $\widehat{x}_{\text{unnorm}} = (\texttt{scale\_in}\sqrt{N})\,\psi'$. To recover the \emph{unitary} FFT (scaled by $1/\sqrt{N}$), we multiply only by \texttt{scale\_in}: $\widehat{x}_{\text{unitary}} = \texttt{scale\_in}\,\psi'$. Other FFT conventions (e.g., $1/N$ on the forward or inverse) are handled by the corresponding one-line rescaling. 
\end{enumerate}

\subsection{Implementation}
\begin{figure}[t]
    \centering
    \includegraphics[width=0.75\textwidth]{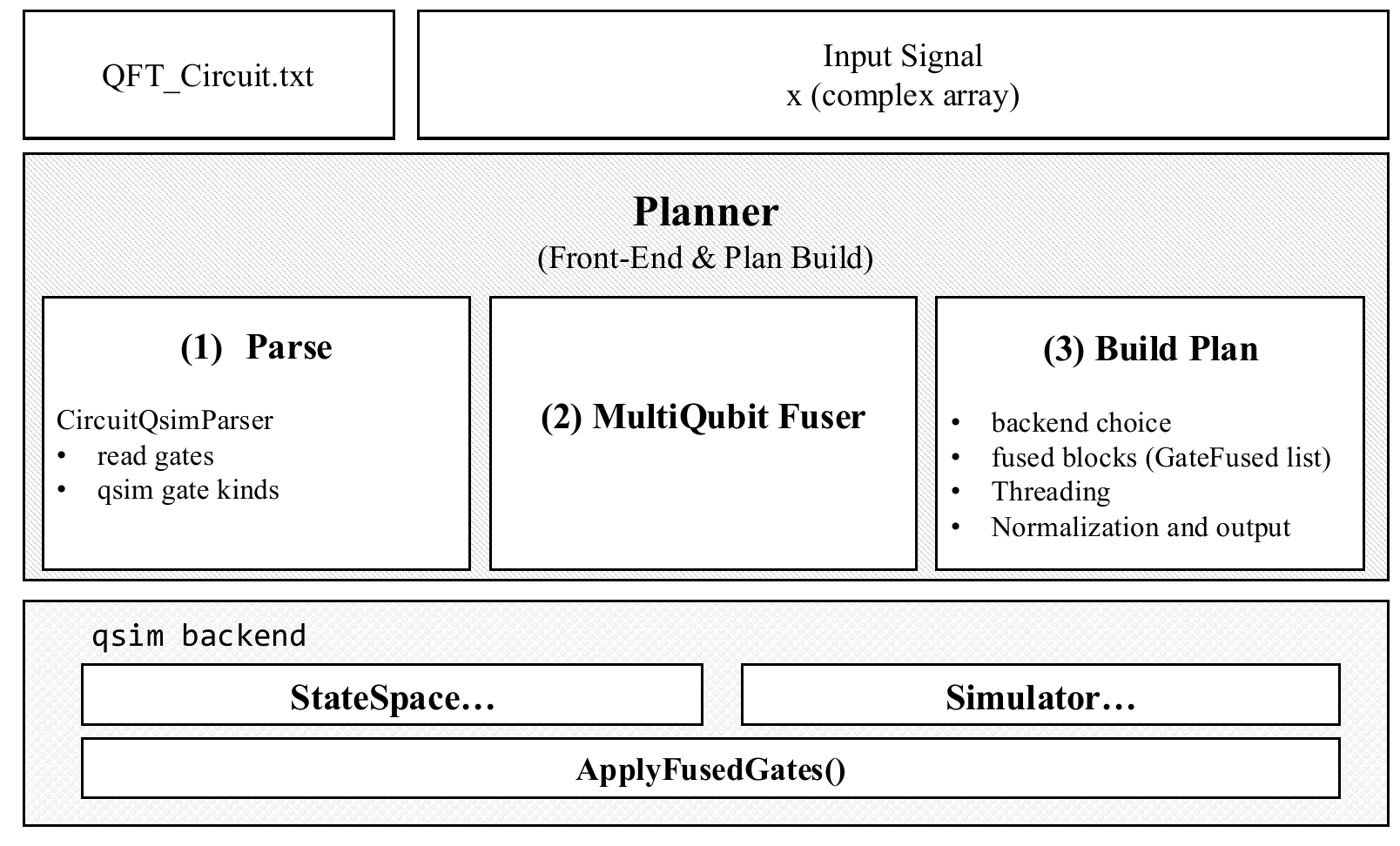}
    \caption{Diagram of the QFT planner.}
    \label{fig:planner}
\end{figure}

Our implementation introduces a planner abstraction that decouples circuit preparation from execution on the \texttt{qsim} backend, as shown in Fig.~\ref{fig:planner}. 
The planner consumes a circuit provided in an input file. It constructs an intermediate representation of gates, performs multi-gate fusion, and emits a device-agnostic execution schedule. The actual simulation, e.g., the allocation of the state vector and application of kernels, is deferred to a back-end factory. \texttt{qsim} provides three factories that realize the same plan on different architectures: a multithreaded CPU backend, a multithreaded AVX backend, and a CUDA backend.

The front end is common to all targets. We parse circuits into \texttt{Circuit}. Fusion is performed by \texttt{qsim}’s multi-qubit fuser. The fuser scans the time-ordered list and greedily coalesces adjacent gates, acting on an overlapping set of qubits. This is subject to a user-defined cap \texttt{max\_fused\_size}. The result is a vector of fused descriptors. Each descriptor records the participating qubits, the contiguous subrange of original gates folded into the block, and a backend-independent payload. The planner preserves global order at the granularity of fused blocks so that downstream execution can process the plan with a single pass.

The QFT plan is backend-agnostic. It contains no device pointers, no launch configurations, and no vector widths. It is defined by the tuple (\texttt{num\_qubits, fused\_blocks, max\_fused\_size}) and aggregate counters of 1- and 2-qubit gates before fusion. This separation allows a single planning pass to feed multiple executors. The factories provide the missing pieces at run time. It provides \texttt{StateSpace} that owns and manages amplitude buffers, and a \texttt{Simulator} that implements the gate kernels over that layout. On the CPU path, the factory binds \texttt{StateSpace} to a flat \texttt{AoS} complex layout and \texttt{Simulator} to a parallel for (implemented with OpenMP) over contiguous index ranges. On the AVX path, the same plan is executed by \texttt{StateSpaceAVX/SimulatorAVX}. This exposes the same abstract vector interface but realizes kernels with wide loads, lane permutations, and fused multiply-adds. On CUDA, \texttt{StateSpaceCUDA} manages device allocation and stream ownership. \texttt{SimulatorCUDA} maps each fused block to one or more kernels whose grid and block dimensions are inferred from the qubit footprint and whether the block is diagonal or dense.

The interaction with core \texttt{qsim} library components follows a determined sequence. The planner uses \texttt{CircuitQsimParser} to construct a \texttt{Circuit}. It then calls the fuser to fuse blocks. No \texttt{StateSpace} or \texttt{Simulator} objects are created during planning. Those are back-end concerns and are instantiated only when a client decides to execute the plan. During execution, the factory produces a zero-initialized \texttt{StateSpace\allowbreak::State} with size $2^n$, initializes it from host input if provided. Then, it iterates the fused blocks: dense blocks are applied through \texttt{ApplyFusedGate(sim, fused[i], state)}, which performs the appropriate tensor contraction on the targeted lanes. Because the fused representation exposes the qubit span, all three executors can make efficient low-level decisions without inspecting individual gates.

\section{Experimental Setup}
We evaluate our QFT→FFT library on single-node platforms using three \texttt{qsim} backends: multithreaded OpenMP, AVX-vectorized CPU kernels, and CUDA. 
For the CPU experiments, we use a node of two AMD EPYC Zen2 2.25\,GHz processors (64 cores each; 128 physical cores total) on the Dardel supercomputer, running SUSE Linux. We compile with \texttt{gcc} using \texttt{-O3} and enable OpenMP for all multithreaded runs. For the GPU experiments, we use the CUDA backend on a system equipped with an NVIDIA A100 GPU, running CentOS~8 (Linux kernel~4.18).  Unless otherwise stated, the CUDA backend uses the default kernel launch parameters provided by \texttt{qsim}. On the CUDA backend, we time kernel execution only using device events around the fused-kernel sequence. The reported performance results exclude host to/from device transfers of input/output arrays, simulator state allocation, and plan construction.

We validate numerical results and compare performance against FFTW on the CPU node. Our baseline is FFTW~v3.3.10.9 built with the \texttt{pthreads} threading interface. We use the single-precision API (\texttt{fftwf\_*}) and enable FFTW’s multithreading for all runs. Plans are created with \texttt{FFTW\_ESTIMATE} to avoid plan-time autotuning and to match our planner’s low setup overhead. All measurements are reported in single precision.

For correctness checks, we use single-tone inputs at multiple sizes. For each target qubit count $n$, we provide an input file of length $2^n$ whose complex samples encode a single discrete complex exponential at a known bin $k_0$. We choose $k_0=2$. In a dense state-vector simulator, such as \texttt{qsim}, the choice of $k_0$ does not alter the kernel’s memory access pattern, which is determined by the circuit and backend layout. The amplitudes are normalized to unit $\ell_2$ norm (to satisfy the simulator’s state constraints). After running the QFT circuit, we rescale the output back to the FFT convention used for reference.

For every configuration (backend, number of qubits, thread count, or AQFT cutoff $k$), we run between 3–5 independent trials and record the circuit simulation execution time. We report the median across trials as the point estimate, and show performance variability with standard deviation from the measurements (shown as error bars in the figures). The same procedure is applied to all experiments and plots.

To assess the cost--accuracy tradeoff of the approximate QFT, we run a sweep over the cutoff $k$ on a 22-wire circuit ($N=2^{22}$). The cutoff limits the number of controlled-phase rotations retained per stage. With a large $k$, we approach the exact QFT (which uses $n(n-1)/2=231$ controlled rotations for $n=22$). Smaller $k$ progressively removes the smallest-angle terms. Tab.~\ref{tab:aqft-k22} reports the resulting gate counts for the diagonal factors, expressed in our notation as controlled-$R_m$ gates with $R_m=\mathrm{diag}(1,e^{-2\pi i/2^{m}})$, and the percentage of their share relative to the exact circuit. 

\begin{table}[t]
\centering
\small
\caption{Controlled-$R_m$ gates in a 22-qubit QFT with approximation cutoff $k$.\label{tab:aqft-k22}}
\begin{tabular}{r @{\hspace{1.6em}} r @{\hspace{1.6em}} r}
\toprule
Cutoff $k$ & \# controlled-$R_m$ gates & \% of exact \\
\midrule
21 (exact) & 231 & 100.0\% \\
16         & 216 & 93.5\%  \\
12         & 186 & 80.5\%  \\
10         & 165 & 71.4\%  \\
8          & 140 & 60.6\%  \\
6          & 111 & 48.1\%  \\
4          &  78 & 33.8\%  \\
\bottomrule
\end{tabular}
\end{table}

We compare each AQFT run (cutoff $k$) to the exact QFT using two standard measures. First, we use \emph{Fidelity} $F$, which is the squared, normalized inner product between the two output vectors. We report $1\!-\!F$ in the plots (lower is better) as an overall similarity error. Second, we use \emph{Top-bin leakage}, which leverages our single-tone input. In this metric, we identify the correct spectral bin from the exact run and measure the fraction of total energy ($Frac_{eng}$) that lands in that bin for each $k$. Leakage is measured as $1 - Frac_{eng}$. In brief, the overall similarity error ($1\!-\!F$) reflects whole-vector agreement, while top-bin leakage ($1 - Frac_{eng}$) quantifies how much energy is spilled out of the intended bin.

\section{Results}
\begin{figure}[t]
\centering

\begin{minipage}{0.48\textwidth}
    \centering
    \includegraphics[width=\imgwidth]{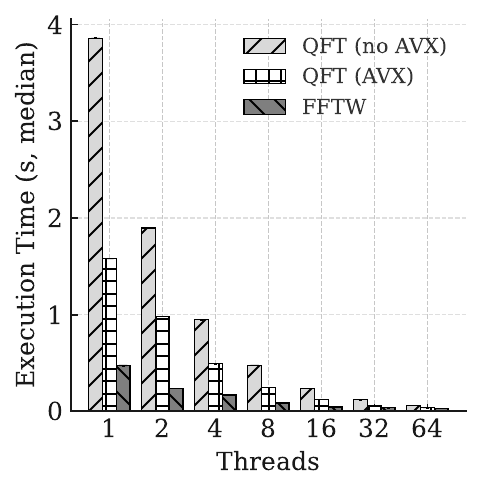}
    \caption{Execution time vs.\ thread count for a 22-qubit input on an AMD EPYC CPU. FFTW is fastest across the sweep, while AVX substantially closes the gap at high thread counts (within $\sim\!16\%$ of FFTW at 64 threads).}
    \label{fig:threads}
\end{minipage}
\hfill
\begin{minipage}{0.48\textwidth}
    \centering
    \includegraphics[width=\imgwidth]{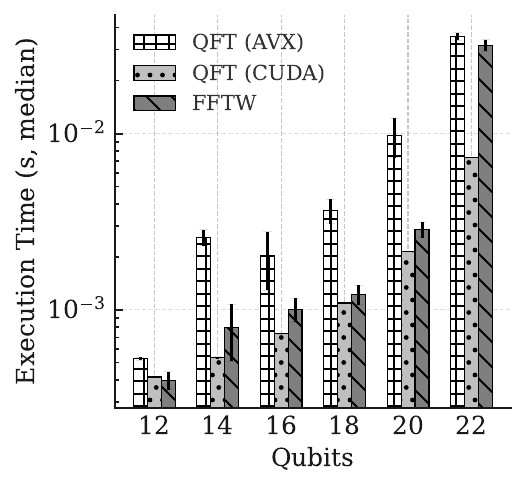}
    \caption{Execution time vs.\ qubits ($n$) for QFT→FFT(AVX), QFT→FFT (CUDA), and FFTW. 
    Times are medians with standard deviation; $y$-axis is log-scale. CUDA is close to FFTW at small $n$, becomes fastest by $n=14$, and is $4.3$--$4.9\times$ faster at $n=22$. CUDA times report kernel execution only (no host–device transfers).}
    \label{fig:cuda_avx}
\end{minipage}
\end{figure}

We first measure CPU performance as a function of thread count on a 22-qubit input ($N=2^{22}$), comparing three paths: QFT→FFT without and with AVX, and FFTW. Fig.~\ref{fig:threads} reports the median execution time with standard deviation. All implementations accelerate with more threads up to 64. QFT (AVX) improves from $\approx 1.60$\,s (1 thread) to $\approx 0.037$\,s (64 threads), a $\sim\!42\times$ speedup; QFT (no-AVX) goes from $\approx 3.86$\,s to $\approx 0.064$\,s ($\sim\!60\times$); FFTW goes from $\approx 0.47$\,s to $\approx 0.032$\,s ($\sim\!15\times$). FFTW is fastest overall, but at high thread counts AVX narrows the gap: at 64 threads, QFT (AVX) is within $\sim\!16\%$ of FFTW ($0.037$\,s vs.\ $0.032$\,s), whereas the non-AVX path remains $\sim\!1.7\times$ slower than AVX. Error bars are negligible.

We next assess the CUDA backend on an A100 and compare it to QFT→FFT (AVX) and FFTW by varying the number of qubits $n\in\{12,\dots,22\}$. This is equivalent to the input array lengths of $N=2^n$, ranging from $2^{12}=4{,}096$ to $2^{22}=4{,}194{,}304$. Fig.~\ref{fig:cuda_avx} shows median execution time with standard deviation on a log-scale $y$-axis. At small sizes, the three backend performances are close (for $n=12$: FFTW $\approx 0.398$\,ms, CUDA $\approx 0.416$\,ms, AVX $\approx 0.527$\,ms). 
CUDA becomes fastest by $n=14$ (CUDA $\approx 0.534$\,ms vs.\ FFTW $\approx 0.793$\,ms). At $n=22$, CUDA is $\approx 7.3$\,ms versus $35.8$\,ms (QFT-AVX) and $31.8$\,ms (FFTW), i.e., about $4.9\times$ faster than AVX and $4.3\times$ faster than FFTW. As $n$ grows, and scaling is driven by memory traffic and parallel throughput, CUDA benefits most and pulls ahead, while AVX narrows but does not eliminate the gap to FFTW at the largest size.

\begin{figure}[t]
\centering
\begin{minipage}{0.48\textwidth}
    \centering
    \includegraphics[width=\imgwidth]{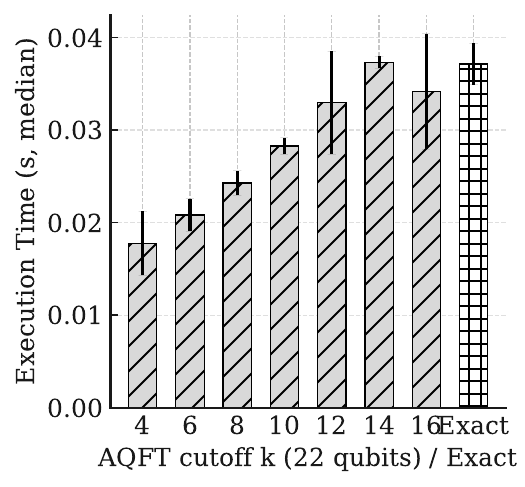}
    \caption{Median execution time for a 22-qubit AQFT vs. cutoff $k$ on the AVX (64-thread) CPU path, with the exact QFT baseline shown for reference. Lower $k$ prunes more controlled-$R_m$ rotations and reduces time, delivering up to $\sim\!2.1\times$ speedup at $k{=}4$.}
    \label{fig:aqft}
\end{minipage}
\hfill
\begin{minipage}{0.48\textwidth}
    \centering
    \includegraphics[width=\imgwidth]{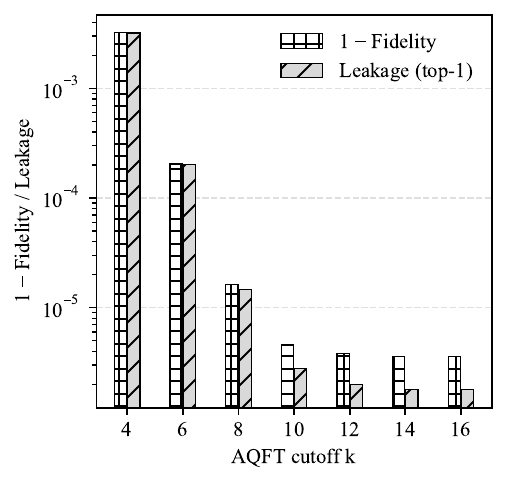}
    \caption{AQFT accuracy for a 22-qubit transform as a function of cutoff $k$. $1\!-\!F$ (one minus fidelity) and top-bin leakage on a log scale (lower is better).}
    \label{fig:error_aqft}
\end{minipage}
\end{figure}

We quantify the effect of the AQFT cutoff on runtime for $n=22$ ($N=2^{22}$) on the AVX path with 64 threads.  Fig.~\ref{fig:aqft} shows median time and standard deviation as $k$ decreases. Pruning small-angle controlled phases reduces work, from $\approx 34.2$\,ms at $k=16$ to $\approx 17.8$\,ms at $k=4$. Using the exact QFT as a reference ($37.1$\,ms), the corresponding speedups are $1.09\times$ ($k{=}16$), $1.13\times$ ($k{=}12$), $1.31\times$ ($k{=}10$), $1.53\times$ ($k{=}8$), $1.79\times$ ($k{=}6$), and $2.09\times$ ($k{=}4$); $k{=}14$ is effectively at parity within measurement noise. 

Finally, we evaluate the accuracy of AQFT on $n=22$ as a function of the cutoff $k$. Fig.~\ref{fig:error_aqft} reports two metrics, $1\!-\!F$ (one minus fidelity) and \emph{top-bin leakage}, both on a log scale for $k\in\{4,6,8,10,12,14,16\}$. Errors decrease with larger $k$. The close tracking of $1\!-\!F$ and top-bin leakage indicates that truncating small controlled phases mainly redistributes energy from the dominant spectral line into nearby side lobes. Restoring deeper rotations re-concentrates that energy.

\section{Related Work}
Work on FFTs spans decades of algorithm design and highly-engineered libraries. \texttt{FFTW} pioneered the planner/plan model and automatic code generation (codelets) to adapt computation to the target machine, to achieve optimal performance across sizes and architectures~\cite{frigo1998fftw}.  \texttt{SPIRAL} extended this idea with algebraic program generation and search-driven optimization across transform families and targets. Vendor libraries such as Intel \texttt{oneMKL}, NVIDIA \texttt{cuFFT} and AMD \texttt{rocFFT} provide production-grade FFTs on GPUs. For a distributed-memory system, among the most famous HPC libraries, there are \texttt{P3DFFT}~\cite{pekurovsky2012p3dfft}, \texttt{AccFFT}~\cite{gholami2015accfft}, and \texttt{heFFTe}~\cite{ayala2020heffte}, which focus on communication-optimal decompositions and heterogeneity. All these works treat the FFT as a classical numerical kernel and improve the performance via algorithmic variants, data movement minimization, and auto-tuning.

The quantum computing community has studied the QFT as a primitive for QPE and factoring, e.g., Shor. Approximate QFTs (AQFTs) truncate small-angle controlled rotations beyond a cutoff, reducing depth with bounded error, an idea dating to Coppersmith~\cite{coppersmith2002approximate} and further developed~\cite{nam2020approximate}.  More recently, structural observations show that, under suitable qubit orders, the QFT exhibits limited entanglement and admits compact classical representations~\cite{chen2023quantum}.

HPC simulation of quantum circuits has become its own HPC area. State-vector simulators, such as Google’s \texttt{qsim} (with backends of SSE/AVX/CUDA/HIP variants and a C++ API), IBM \texttt{Qiskit Aer}, and NVIDIA’s \texttt{cuQuantum/\allowbreak cuStateVec}~\cite{bayraktar2023cuquantum}, provide multithreaded and GPU-accelerated kernels for single- and multi-qubit gate application, measurement, and related operations. The \texttt{QuEST} project demonstrated strong single-node and distributed scaling with CPU+GPU backends~\cite{jones2019quest}. In addition, Intel’s Intel Quantum Simulator (\texttt{IQS})~\cite{guerreschi2020intel}, formerly \texttt{qHiPSTER}, is a distributed-memory state-vector engine optimized for multi-core and multi-node machines, with vectorized kernels and MPI support. \texttt{PennyLane} provides a family of devices spanning research prototyping to HPC: \texttt{default.qubit} and HPC \texttt{lightning.qubit} and \texttt{lightning.gpu}~\cite{asadi2024hybrid}. These systems aim to accelerate quantum algorithm prototyping, but their primitives align with the performance concerns of classical HPC kernels.  


\section{Discussion and Conclusion}
Recent progress in quantum computing is providing ideas that could benefit HPC beyond quantum computers. This work applies those ideas to the Fourier transform. We introduce QFT→FFT, a class of FFT libraries that realize the DFT by executing a QFT circuit on classical state-vector simulators. 

Beyond the QFT, this approach can, in principle, address other core HPC tasks such as eigenvalue estimation. In particular, one can simulate QPE~\cite{nielsen2010quantum} on a classical state-vector simulator. This would consist of preparing a state with overlap on the eigenvectors of an operator of interest, applying a short sequence of controlled time-evolutions of that operator, and finally using a Fourier-style readout to convert the observed phases into eigenvalue estimates~\cite{cruz2020optimizing}. Conceptually, QPE is just a sequence of small matrix operations on the state vector, akin to our QFT runs. However, it is arranged to reveal spectral information rather than a transform. This simulator-based QPE might be an appealing tool for exploring eigenproblems that arise in scientific computing. 

Despite higher asymptotic cost on classical hardware, $O(N\log^2 (N))$, for exact QFT vs. $O(N\log N)$ for FFT, our QFT→FFT design is competitive in practice. On AMD EPYC Zen2, the AVX backend matches multithreaded FFTW at large input size. Two mechanisms close the gap: planner-driven gate fusion and layout adapters that realize stride permutations as low-cost I/O transforms tuned to each backend. Truncation of small-angle rotations (AQFT with cutoff $k$) reduces simulator cost to $O(Nk\log N)$. For a constant $k$, this matches FFT complexity and yields additional speedups on both CPU and GPU while preserving good accuracy. These results indicate that QFT→FFT is a viable and portable alternative to classical FFT libraries. 


Our QFT→FFT library uses a state-vector simulator, but other simulation paradigms can offer better performance in specific regimes. For instance, the tensor-network simulator represents the $n$-wire state as a chain or tree of low-rank tensors~\cite{nguyen2022tensor} (e.g., Matrix Product States and Matrix Product Operators) and contracts small gate tensors into them. When intermediate entanglement is limited or localized~\cite{chen2023quantum}, the bond dimension remains moderate, so both time and memory scale with the bond dimension rather than $2^n$~\cite{berezutskii2025tensor}. The QFT circuit is shallow and diagonal between Hadamards, and the AQFT further removes long-range small-angle couplings. These properties make QFT/AQFT amenable to tensor-network contraction, especially with input vectors that are structured or compressible. One can contract gates in stage order, use slicing over a few wires to cap memory, and map contractions to GPUs. A potential other direction is Monte Carlo simulation in the \emph{sum-over-histories} style \cite{rudiak2006sum}. The amplitude for output index $y$ is a complex sum of contributions with phases determined by the controlled rotations. Sampling input indices (or short computational paths) with importance weights leads to unbiased estimators of $|\langle y|\mathrm{QFT}\,|x\rangle|$. This Monte Carlo approach for QFT is a valuable technique when only magnitudes or the power spectrum are required. In addition, Monte Carlo variance reduction can be applied to this approach. All these alternatives fit our planner/plan interface. The planner selects a backend (state vector, tensor network, or Monte Carlo) per size and accuracy budget, applies the same fusion opportunities, and exposes the same layout adapters. Exploring such backends might lead to further speedups and enables much larger problem sizes than a pure state-vector approach, as presented in this paper.

In summary, QFT→FFT reformulates FFT as a gate-execution problem and, combined with planner-driven fusion and AQFT truncation, delivers performance comparable to state-of-the-art FFT libraries. Matching multithreaded FFTW on CPU is a state-of-the-art milestone for classical QFT simulation. This indicates that a circuit-execution pathway can rival mature FFT libraries. This parity, combined with AQFT’s controllable accuracy-performance trade-offs, positions QFT→FFT as a practical alternative in HPC. We expect this approach to scale with distributed simulators and alternative backends (tensor networks, Monte Carlo), enabling larger problems and further performance improvement.

\bibliographystyle{ACM-Reference-Format}
\bibliography{qft_arxiv}

\end{document}